\documentclass[APS,twocolumn,preprintnumbers]{revtex4}

\usepackage{amsmath,amsfonts,amssymb}
\usepackage{pstricks}
\usepackage{bbm}
\usepackage{graphicx}

\newcommand{\ie}{i.e.\ }

\newcommand{\etc}{etc.}

\newcommand{\bra}[1]{\langle #1|\,}
\newcommand{\ket}[1]{\,|#1 \rangle}
\newcommand{\braket}[2]{\langle #1|#2 \rangle}
\newcommand{\ketbra}[2]{\ket{#1}\negthinspace\bra{#2}}
\newcommand{\mypict}[1]{\,\includegraphics[height=1.5em]{th_#1.jpg}}

\begin{document}

\title{Quantum complex networks}
\author{S. Perseguers$^1$, M. Lewenstein$^{2,3}$, A. Ac\'in$^{2,3}$ \& J. I. Cirac$^1$}
\affiliation{
    $^1$Max-Planck--Institut f\"ur Quantenoptik, Hans-Kopfermann-Strasse 1, 85748 Garching, Germany\\
    $^2$ICFO--Institut de Ci\`encies Fot\`oniques, Mediterranean Technology Park, 08860 Castelldefels, Spain\\
    $^3$ICREA--Instituci\'o Catalana de Recerca i Estudis Avan\c cats, Lluis Companys 23, 08010 Barcelona, Spain
}
\date{\today}

\begin{abstract}
In recent years, new algorithms and cryptographic protocols based on the laws of
quantum physics have been designed to outperform classical communication and computation.
We show that the quantum world also opens up new perspectives in the field of complex networks.
Already the simplest model of a classical random network changes dramatically
when extended to the quantum case, as we obtain a completely distinct behavior
of the critical probabilities at which different subgraphs appear. In particular,
in a network of $N$ nodes, any quantum subgraph can be generated by local
operations and classical communication if the entanglement between pairs of nodes
scales as $N^{-2}$.
\end{abstract}

\maketitle

On the one hand, \textit{complex networks} describe a wide variety of systems in nature and
society, as chemical reactions in a cell, the spreading of
diseases in populations or communications using the
Internet \cite{AB02}. Their study has traditionally been the
territory of graph theory, which initially focused on regular
graphs, and was extended to random graphs by the
mathematicians Paul Erd\H{o}s and Alfr\'ed R\'enyi in a series of
seminal papers \cite{ER59,ER60,ER61} in the 1950s and 1960s.
With the improvement of computing power and the emergence of
large databases, these theoretical models have
become increasingly important, and in the past few years new
properties which seem universal in real networks have been described,
as a small-world \cite{WS98} or a scale-free \cite{BA99} behavior.

On the other hand, \textit{quantum networks} are expected to be
developed in a near future in order to achieve, for instance, perfectly secure
communications \cite{K08,GT07}. These networks are based on the
laws of quantum physics and will offer us new
opportunities and phenomena as compared to their classical
counterpart. Recently it has been shown that quantum
phase transitions may occur in the entanglement properties of
quantum networks defined on regular lattices, and that the use of joint
strategies may be beneficial, for example, for quantum teleportation between
nodes \cite{ACL07,PCA+08}. In this work we introduce a simple model
of {\it complex quantum networks}, a new class of systems
that exhibit some totally unexpected properties. In fact we
obtain a completely different classification of their behavior as
compared to what one would expect from their classical counterpart.\\

\begin{figure}
\begin{center}
    \includegraphics[width=.9\linewidth]{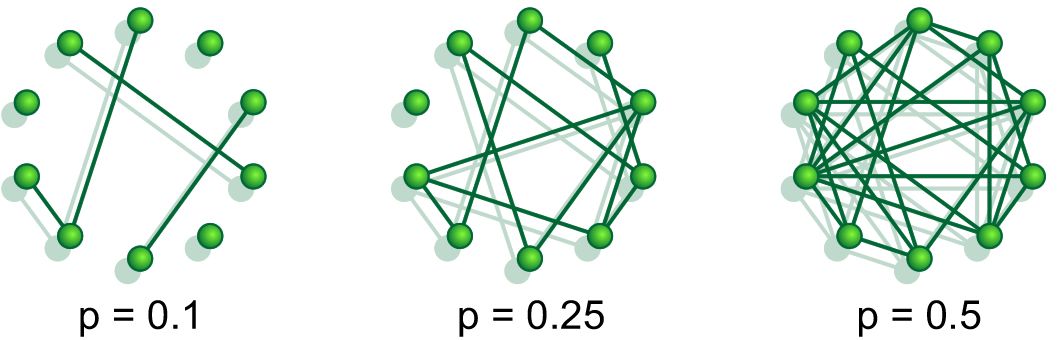}
    \caption{Evolution process of a classical random graph with $N=10$ nodes: starting
    from isolated nodes, we randomly add edges with increasing probability
    $p$, to eventually get the complete graph $K_{10}$ for $p=1$.}
    \label{fig:cRG}
\end{center}
\end{figure}

A classical network is mathematically represented by a graph,
which is a pair of sets $G=(V,E)$ where $V$ is a set of $N$ nodes
(or vertices) and $E$ is a set of $L$ edges (or links) connecting two
nodes. The theory of random graphs, aiming to tackle networks with
a complex topology, considers graphs in which each pair of nodes
$i$ and $j$ are joined by a link with probability $p_{i,j}$.
The simplest and most studied model is the one where this
probability is independent of the nodes, with $p_{i,j}=p$, and the
resulting graph is denoted $G_{N,p}$. The construction of these
graphs can be considered as an evolution process: starting
from $N$ isolated nodes, random edges are successively added and
the obtained graphs correspond to larger and larger connection
probability, see Fig.~\ref{fig:cRG}. One of the main goals of random-graph
theory is to determine at which probability $p$ a specific
property $P$ of a graph $G_{N,p}$ mostly arises, as $N$ tends to
infinity. Many properties of interest appear suddenly, \ie there
exists a critical probability $p_c(N)$ such that almost every
graph has the property $P$ if $p\geq p_c(N)$ and fails to have it
otherwise; such a graph is said to by \textit{typical}. For
instance, the critical probability for the appearance of a given
subgraph $F$ of $n$ nodes and $l$ edges in a typical random graph
is \cite{B85}:
\begin{equation}
    p_c(N)=c\,N^{-n/l},
\end{equation}
with $c$ independent of $N$. It is instructive to look at the
appearance of subgraphs assuming that
$p(N)$ scales as $N^z$, with $z\in(-\infty,0]$ a tunable
parameter: as $z$ increases, more and more complex subgraphs
emerge, see Tab.~\ref{tab:classical-pc}.\\

\begin{table}
\begin{center}
    \begin{tabular}{c|@{\quad}c@{\qquad}c@{\qquad}c@{\qquad}c@{\qquad}c@{\qquad}c@{\quad}}
        \hline\hline
        $\,z\,\vphantom{\frac{|^|}{|_|}}$ & $-\infty$& $-2$ & $-\frac{3}{2}$ & $-\frac{4}{3}$ &
        $-1$ & $- \frac{2}{3}$\\
        \hline
        &&&&&&\\[-1.1em] &
        \mypict{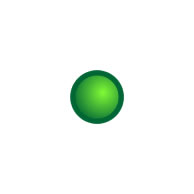} & \mypict{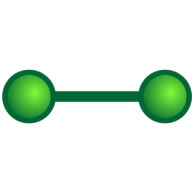} & \mypict{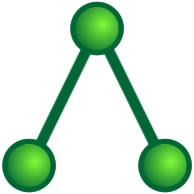} & \mypict{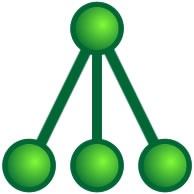} &
        \mypict{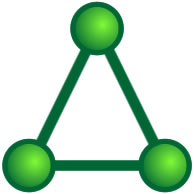} \mypict{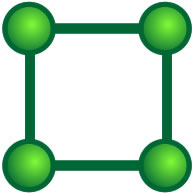} & \mypict{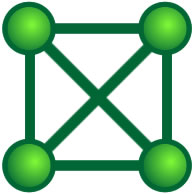} \\
        \hline\hline
    \end{tabular}
    \caption{Some critical probabilities at which a given subgraph $F$ appears in
    	classical random graphs of $N$ nodes connected with probability $p\sim N^z$:
        cycles and trees of all orders appear at $z=-1$, while complete subgraphs
        (of order four or more) appear at a higher connection probability \cite{AB02}.}
    \label{tab:classical-pc}
\end{center}
\end{table}

\begin{figure}
\begin{center}
    \includegraphics[width=.9\linewidth]{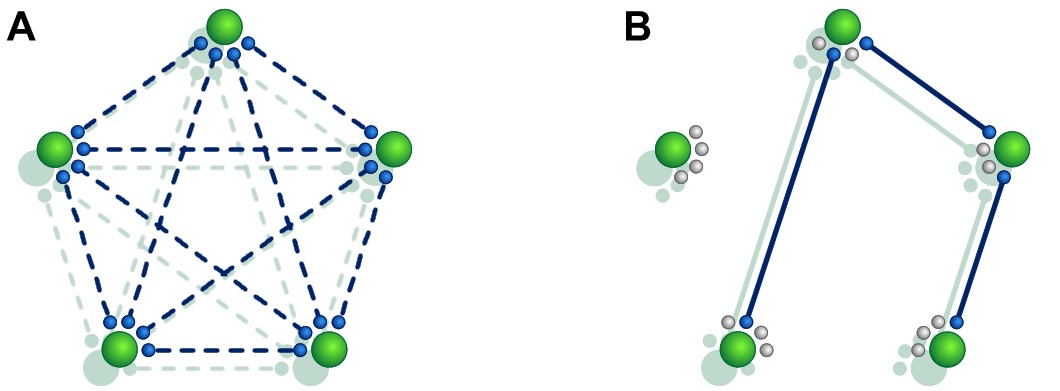}
    \caption{An example of a quantum random graph
    on five nodes. \textbf{(A)} Each node is in possession of four qubits which are entangled with qubits
    belonging to the other nodes. All the connections are identical and pure
    but non-maximally entangled pairs. \textbf{(B)} Imperfect pairs can be converted
    into maximally entangled ones with some probability of success $p$ (here
    $p=0.25$); this strategy mimics the behavior of classical random graphs.}
    \label{fig:qRG}
\end{center}
\end{figure}

We consider now the natural extension of the previous scenario to
a quantum context. For each pair of nodes we replace the
probability $p_{i,j}$ by a quantum state $\rho_{i,j}$ of two
qubits, one at each node. Hence every node possesses $N-1$ qubits
which are pairwise entangled with the qubits of the other nodes,
as depicted in Fig.~\ref{fig:qRG}. The goal is then to establish maximally entangled states
within certain subsets of nodes by using protocols that consist of
local operations and classical communication (LOCC) \cite{W89}, where each
node can apply different measurements on its quantum system and then
communicates its results to the rest. This is in fact a very natural
scenario since maximally entangled states are the resource for
most quantum information tasks, and thus the goal of any strategy.
As in the classical random graphs we
consider that pairs of qubits are identically connected, with
$\rho_{i,j}=\rho$. Furthermore, we restrict ourselves to the
simplest case where the states are pure, \ie
$\rho=\ketbra{\varphi}{\varphi}$, since it already leads to some
very intriguing phenomena. We take these states to be:
\begin{equation}
    \ket{\varphi} = \sqrt{1-p/2}\ket{00} + \sqrt{p/2}\ket{11},
    \label{eqn:intro-phi}
\end{equation}
where $0\le p\le 1$ measures the degree of entanglement of the
links. As before $p$ scales with $N$ and we write $\ket{G_{N,p}}$
the corresponding quantum random graph. The choice of the coefficients
in Eq.~(\ref{eqn:intro-phi}) becomes clear if one considers
the following simple strategy where links are treated
independently: each pair of qubits tries to convert its connection by LOCC
into the maximally entangled state $\ket{\Phi^+}\propto\ket{00}+\ket{11}$.
The probability of a successful conversion is $p$, which is optimal \cite{N99,V99},
and therefore the task of determining the type of maximally
entangled states remaining after these conversions can be exactly
mapped to the classical problem. In that case we obtain the results of
Tab.~\ref{tab:classical-pc}, and for example for $z\ge -2$ the
probability to find a pair of nodes that share a maximally entangled
state is one, whereas that of having three nodes sharing three
maximally entangled states is zero unless $z\ge -1$.\\

\begin{figure}
\begin{center}
    \includegraphics[width=.9\linewidth]{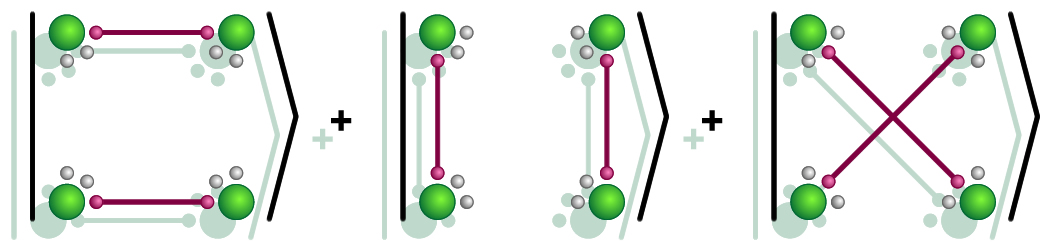}
    \caption{Graphical representation of $\ket{K_c}$.
    Each node has exactly $c-1$ orthogonal states in which $c-2$ qubits
    are $\ket{0}$ and one is $\ket{1}$;
    qubits that were originally entangled share the same state. A link
    denotes, here, the separable state $\ket{11}$ while $\ket{00}$ is not drawn, and
    $\ket{K_c}$ is defined as the coherent superposition of all perfect matchings of the
    complete graph $K_c$. In this example an appropriate labeling of the states
    leads to $\ket{K_4} \propto \ket{1111} + \ket{2222}+\ket{3333}$.}
    \label{fig:GHZ4}
\end{center}
\end{figure}

Allowing strategies which entangle the qubits within
the nodes offers new possibilities and brings powerful results. This is indeed
a general fact in quantum information theory as, for
example, in the context of distillation of entanglement \cite{BBP+96}
or in the non-additivity of the classical and quantum capacity of quantum channels \cite{H09}.
As a first illustration of the advantage of joint actions on
qubits we show how some relevant multipartite entangled states, like
the three-dimensional Greenberger-Horne-Zeilinger (GHZ) \cite{GHZ89}
state of four particles, can be obtained when $z=-2$.
At each node we apply a generalized (or incomplete)
measurement \cite{NC00} whose elements $P_m$ are projectors onto
the subspaces consisting of exactly $m$ qubits in the state
$\ket{1}$, \ie we count the number $m$ of ``links''
$\ket{11}$ attached to the nodes without revealing their precise
location. Remark that such links are separable and so do not
represent the sought connections $\ket{\Phi^+}$. For each measurement we
get a random outcome $m$ whose value is either $0$ or $1$ since,
as in the classical case, the probability to get $m\ge 2$ is zero
for infinite $N$ \cite{B81}, see App.~A. Setting $p=2cN^{-2}$
one finds that there are in average $c$ nodes having exactly one qubit
in the state $\ket{1}$, whereas all nodes where $m=0$ factor out since
they are completely uncorrelated with the rest of the system. In
Fig.~\ref{fig:GHZ4} we describe the remaining state, which we write $\ket{K_c}$,
and show that for $c=4$ it is a GHZ state of four qutrits.\\

We now turn to the main result of this paper: for $N$
tending to infinity and with $p\sim N^{-2}$ one is able to obtain with unit
probability a quantum state with the structure of any finite
subgraph. That is, for an arbitrary subgraph $F=(V,E)$ composed of $n$
vertices and $l$ edges, one can get the state $\ket{F}$
consisting of $l$ maximally entangled pairs $\ket{\Phi^+}$ shared
among $n$ nodes, according to $E$:
\begin{equation}
    \ket{F} = \bigotimes_{i=1}^l \ket{\Phi^+}_{E_i}.
    \label{eqn:F}
\end{equation}
This result implies that in the quantum case the structure of
Tab.~\ref{tab:classical-pc} completely changes, as all subgraphs
already appear at $z=-2$. Actually, our results are more general since
any state of some interest in quantum information theory, as W
states, Dicke states, graph states, \etc, also arise
at $z=-2$. We sketch here the four steps of our proof and refer the reader
to App.~B for detailed calculations.
First, we create the state $\ket{K_c}$ described in Fig.~\ref{fig:GHZ4}, with $c=n+D$, $D=d^2$ and $d=2^l$: $n$ nodes
will be kept to build the final state while $D$ additional nodes
are needed to establish the desired quantum correlations.
Second, we measure all links corresponding to the subgraph $K_n$ and
the other $D$ vertices are measured in the Fourier basis, which leads
to a highly entangled state of $n$ nodes of dimension $D$.
Third, in order to extract the right correlations, we split each
node into two subsystems of dimension $d$, measure one of them in
the Fourier basis and post-select a specific outcome pattern.
Finally, we show that the resulting state, a $n$-partite GHZ state
of dimension $d$, can be converted into $\ket{F}$ by some appropriate projections.\\

At this point let us briefly describe a possible setup for
an implementation of our ideas, where atoms store the
quantum information, and thus represent qubits, while photons
are used to create remote entanglement. This scenario is
currently well admitted to be the most promising one for a realization of
quantum networks \cite{K08}. In particular, we consider the case
where (continuous-variable) entanglement contained in a two-mode squeezed
light is transformed into (discrete) entanglement between atoms trapped
in distant high-quality cavities \cite{KC04}. Assuming perfect operations one can
drive the system so that its steady state is exactly described by
Eq.~\ref{eqn:intro-phi}, with $p<1$ due to a finite squeezing of the light.
In App.~C we discuss a simple model of mixed-state networks
where the source of squeezed states fails to emit, with some probability, any light.
We show that the quantum phenomena considered so far persist despite these imperfections,
and we believe that this will still be the case for more general types of errors.\\

In conclusion, we have introduced a model of complex quantum
networks based on the theory of random graphs and have shown that
allowing joint actions on the nodes dramatically changes one of
their main properties, namely the appearance of subgraphs
according to the connection probability. In fact, all classical
exponents collapse onto the value $z=-2$ in the quantum case and
we expect a large variety of new phenomena in, for instance,
quantum models of small-world or scale-free networks. The model we
have introduced is quite simple in the sense that
connections are represented by pure states, whereas more realistic
setups involve mixed states. We discussed this problem for a specific type of imperfections,
and even if it may be much more difficult to tackle it in full generality
we are confident of the persistence of unexpected and intriguing phenomena
in real quantum networks. We hope that our results will inspire work in
this direction and shed some light on the very active domain of
complex networks.

\begin{acknowledgments}
	We acknowledge support from the EU projects ``Scala'' and ``QAP", the ERC
    grant ``PERCENT", the DPG excellence cluster ``Munich Center for Advanced Photonics''
    and FOR 635, the QCCC program of the Elite Network of Bavaria, the Spanish MEC
    Consolider QOIT and FIS2007-60182 projects, la Generalitat de Catalunya and Caixa Manresa.
\end{acknowledgments}

\appendix


\subsection*{Appendix A: General considerations and notation}
As in the classical case we consider the exponent $z$ in $p=N^z$
to be larger than or equal to $-2$, since the
overlap of the quantum random graph and the product state $\ket{\vec{0}}$ of all
qubits in $\ket{0}$ approaches unity for $z<-2$ and in the limit of infinite lattices:
\[
    |\braket{G_{N,p}}{\vec{0}}|^2 = \left(1-\frac{p}{2}\right)^{\frac{N(N-1)}{2}}
    \simeq \exp\Big(\frac{-N^{z+2}}{4}\Big)\rightarrow 1,
\]
and therefore no local quantum operation is able to create
entanglement between nodes in this case. Furthermore, the outcome
distribution of the measurements $P_m$ follows the classical
degree distribution of a random graph $G_{N,p'}$, with $p'=p/2$.
In fact the only information we get from $\ket{G_{N,p}}$ by
applying $P_m$ is the number of links $\ket{11}$ attached to a given
node, and each such link gets a weight $p/2$ in the outcome
probabilities. This distribution has been shown to be well
approximated by the one that we get if the nodes are considered to
be independent, so that the expected number $x_m$ of
outcomes $m$ reads, setting $p=2cN^z$ and for $z<-1$:
\begin{align}
    E(x_m) &= \frac{N!}{m!\,(N-1-m)!} \left(\frac{p}{2}\right)^m
    		\left(1-\frac{p}{2}\right)^{N-1-m}\nonumber\\
           &\simeq \frac{c^m}{m!} N^{m(z+1)+1}.
    \label{eqn:expect-order-outcomes}
\end{align}
In the text we often mention the act of measuring a node, or more
generally of a system of dimension $d$: if no basis is specified
we mean a measurement in the computational basis, \ie we project
the system onto the states $\ket{1},\ldots,\ket{d}$. We also use
the expression ``to measure a link'' to indicate that one of its
qubits is measured in the $\{\ket{0},\ket{1}\}$ basis. Finally, we
introduce the notation
\begin{equation}
    \ket{\Phi_n^{k,d}}= \frac{1}{\sqrt{d}}\sum_{j=1}^d e^{\frac{2\pi i}{d}jk}
    \ket{\underbrace{jj\ldots j}_{n}},\quad k\in\{1,\ldots,d\},
\end{equation}
for the Fourier transform of $\ket{\Phi_n^d}\equiv\ket{\Phi_n^{d,d}}$.
The latter state is referred in the text as a GHZ state of
dimension $d$ (on $n$ nodes), and a measurement in the Fourier
basis of a system of dimension $d$ is its projection onto the
states $\ket{\Phi_1^{k,d}}$.

\subsection*{Appendix B: Proof of the main result}

We give here the details of the construction of a state $\ket{F}$,
starting from $\ket{G_{N,p}}$ and using LOCC only. But let us
first note that if one is able to find a construction which
succeeds with a strictly positive probability, say $p_F$, that
does not depend on $N$ but on the subgraph $F$ only, then
$\ket{F}$ can be obtained with a probability arbitrary close to
one. The reason is that we can always subdivide the $N$ nodes into
$L$ sets of $N/L$ nodes, with $L\gg 1/p_F$, and apply the same
construction on each set. These sets can be treated as
independent if we initially discard all links connecting
different sets, which is done by measuring the corresponding qubits.
Note that in what follows we do not try to optimize the procedure
since in the limit of infinite $N$ we are mainly interested in the existence
of a finite probability $p_F$, not in its maximal possible value.\\

{\it First step.}
We start the construction by creating the state $\ket{K_c}$, with $c=n+D$,
$D=d^2$ and $d=2^l$, which can be obtained with a probability
approaching unity in the limit of infinite $N$. In fact one can choose
a value $c'\gg c$, apply the projections $P_m$ on the nodes of the
quantum network and get a system of average size $c'$. The
number of nodes of the resulting state can then be decreased in a deterministic fashion:
one measures all qubits of a node, gets $0$ for all outcomes except for a
random one whose neighbor automatically factors
out. Hence, the system is projected onto $\ket{K_{c'-2}}$ and
the procedure can be iterated until we get $\ket{K_c}$.\\

{\it Second step.}
We remove all connections shared between $n$ nodes of
$\ket{K_c}$, \ie we measure the $n(n-1)/2$ concerned links, and
the operation is successful if all outcomes are 0. In that way
we build a state that is the coherent superposition, as described
in Fig.~\ref{fig:GHZ4}, of all perfect matchings of the join graph of $K_D$
and of the empty graph on $n$ nodes. We further
measure these $D$ nodes, but this time in the Fourier basis
$\{\ket{\Phi_1^{k,c-1}},1\leq k \leq c-1\}$ in order not to
reveal where the links $\ket{11}$ lie. We can correct the possibly
introduced phases and the resulting state reads:
\begin{equation}
    \ket{\varphi_n^D} \propto \sum_{i_1\neq i_2\ldots \neq i_n=1}^D\ket{i_1i_2\ldots i_n}.
    \label{eqn:ketD}
\end{equation}

{\it Third step.}
It is not convenient to deal with sums whose indices are subject
to constraints, so we develop Eq.~\ref{eqn:ketD} to let the sums
freely run from $1$ to $D$. For example, the state on three nodes
is expressed as:
\[
    \ket{\varphi_3^D} = \sum_{i,j,k=1}^D \ket{ijk} - \sum_{i,j=1}^D \big( \ket{iij}
    + \ket{iji} + \ket{jii}\big) + 2 \sum_{i=1}^D \ket{iii}.
\]
More generally this leads to a weighted and symmetric superposition of states
of the form $\bigotimes_{i=1}^{r} \ket{\Phi_{\lambda_i}^D}$ for all partitions
$(\lambda_1,\lambda_2,\ldots,\lambda_r)$ of $n$, where $\sum_{i=1}^r\lambda_i=n$
(a partition of a positive integer is a way of writing it as a sum of positive 
integers). We want to remove all terms of this sum but the
last one, which is a GHZ state that will allow us to obtain
$\ket{F}$. To that purpose we use the fact that $\ket{\Phi_n^D} =
\ket{\Phi_n^{d^2}} = \ket{\Phi_n^d}^{\otimes2}$ for all $n$ and
$d$, split each node into two subsystems of dimension $d$ and
measure one of them in the Fourier basis. This operation is successful if all outcomes are $k=1$,
except the last one which should be $k=d-n+1$. In fact, to see what happens
to a state $\ket{\Phi_m^d}$ we note that $\braket{\Phi_1^{k,d}}
{\Phi_m^{j,d}} \propto \ket{\Phi_{m-1}^{j-k,d}}$ if $m>1$ and
$\delta_{j,k}$ otherwise. Therefore, by sequentially measuring
$\bra{\Phi_1^{1,d}}$, a state $\ket{\Phi_m^d}$ shared among any
$m<n$ nodes transforms as:
\[
    \ket{\Phi_m^d} \equiv \ket{\Phi_m^{d,d}} \mapsto \ket{\Phi_{m-1}^{d-1,d}}
    \mapsto \ldots \mapsto \ket{\Phi_1^{d-m+1,d}} \mapsto 0,
\]
as long as $d>m$. But this is always the case because, without loss of
generality, the subgraph can be considered to be connected, so that $l\geq n-1$
and thus $d\geq2^l\geq2^{n-1}\geq n>m$. Hence all terms vanish except
the GHZ state $\ket{\Phi_n^d}$.\\

{\it Fourth step.}
The last step consists in transforming $\ket{\Phi_n^d}$ into $\ket{F}$. To that
end first expand Eq.~\ref{eqn:F}, \ie write explicitly all its terms in the
computational basis, and group the qubits according to the connections $E$. This
leads to a sum of product states of the form $\ket{\varphi_{i,1}}\ldots\ket{\varphi_{i,n}}$
with $i=1,\ldots,2^l$. Since we have chosen $d=2^l$, we can now apply the measurement
element $\sum_{i=1}^d \ketbra{\varphi_{i,j}}{i}$ on each node $j$ of $\ket{\Phi_n^d}$,
which achieves the desired transformation and concludes the proof.

\subsection*{Appendix C: A mixed-state scenario}
\begin{figure}
\begin{center}
    \includegraphics[width=.9\linewidth]{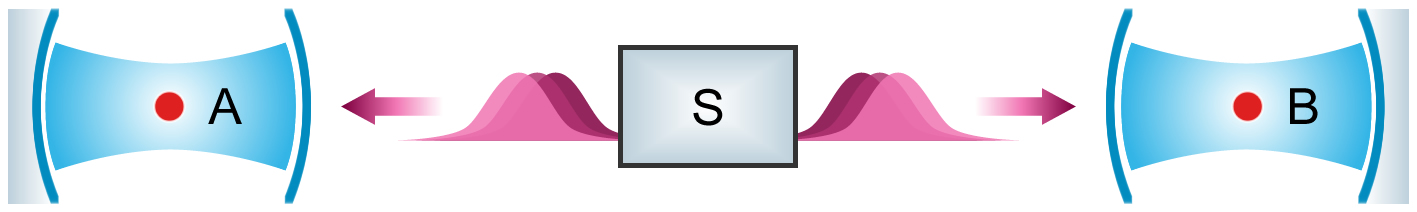}
    \caption{Scheme for entanglement generation between two nodes:
		two distant cavities are simultaneously driven by a common source $S$ of squeezed light,
    	and in the steady state of the system the two atoms $A$ and $B$ become entangled.}
    \label{fig:setup}
\end{center}
\end{figure}
We consider here the setup described in \cite{KC04}, where two high-quality cavities
are simultaneously driven by a common source of two-mode squeezed light, see Fig.~\ref{fig:setup}.
The continuous-variable entanglement contained in the light is transformed into discrete
entanglement between atoms, and the steady state of the system is described by the
non-maximally entangled pure state we use throughout the text. Let us now introduce some
errors into the networks, considering an imperfect source of light which fails to emit squeezed states
with some probability $\varepsilon$. Equivalently, this imperfect source produces the vacuum
state $\ket{00}$ with probability $\varepsilon$ so that the connections of the quantum network are:
\begin{equation}
	\rho = (1-\varepsilon) \ketbra{\varphi}{\varphi} + \varepsilon \ketbra{00}{00},
\end{equation}
with $\ket{\varphi}$ defined in Eq.~\ref{eqn:intro-phi}.
For this error we can show that it is still possible to construct, in the regime $z=-2$,
mixed states which are close to arbitrary quantum subgraphs. To that end let us
go through the four steps of the construction described in the previous paragraph.
The first step consists in creating the state $\ket{K_c}$ by applying the measurements
$P_m$ on the nodes of the network. These measurements are not affected by the imperfect source
since no extra links $\ket{11}$ are added: only the number of outcomes $m=1$
slightly decreases from $c$ to $(1-\varepsilon)c$, but one can choose a larger constant
$c'$ in $p=2c'N^{-2}$ in order to get with certainty $c$ outcomes $P_1$. For $c=4$
for instance, the remaining nodes are in a mixture of the desired state $\ket{K_4}$ and some completely
separable states:
\begin{equation}
	\rho_{K4} = x \ketbra{K_4}{K_4} + \frac{1-x}{3}\sum_{i=1}^3\ketbra{iiii}{iiii},
\end{equation}
with $x=(1-\varepsilon)^2$ for infinite $N$. Despite the presence of separable states,
$\rho_{K4}$ is useful for quantum information tasks since it is distillable for
all $\varepsilon<1$, \ie the coefficient $x$ can be brought arbitrarily close to unity
if one possesses a large number of copies of it. However, in the regime $z=-2$
it is impossible to get several copies of $\rho_{K4}$ on the \textit{same} four nodes.
But this is not a problem since, alternatively, we can repeat the construction $1/x$
times so that any use of $\ket{K_c}$ is still achieved with high probability.
More generally, the state $\rho_{Kc}$
on $c$ nodes is a mixture of $\ket{K_c}$ and of some partially separable states,
with $x$ equal to $(1-\varepsilon)^{c/2}$. Note that all terms appearing
in $\rho_{Kc}$ are also present in $\ket{K_c}$, but they now are probabilistically
weighted. This structure is maintained throughout the construction of the quantum
subgraphs (steps two to four of the proof),
so that with probability $x$ we create $\ket{F}$, and with probability $1-x$
we get a sum of useless quantum states. Therefore, for all $\varepsilon<1$, 
with a strictly positive probability we can achieve quantum communication or computation that
is impossible within a classical framework.

In conclusion, we have discussed a possible implementation of our model of
quantum networks and shown that the proposed construction of quantum subgraphs is robust against
some specific imperfections, namely a noise $\ketbra{00}{00}$ from the light sources.
Remark that this is not the case for errors involving terms like $\ketbra{01}{01}$,
$\ketbra{10}{10}$ or $\ketbra{11}{11}$ in the connections. However, in that case, we are
confident that other quantum strategies (based on purification methods for instance)
will still lead to intriguing and powerful phenomena, thus stimulating further
work on complex quantum networks.

\bibliography{article}

\end{document}